\documentclass[fp,twocolumn]{jpsj3}
\usepackage{siunitx}
\usepackage{txfonts}

\newif\iffigure
\figurefalse
\figuretrue

\title{Successive Antiferromagnetic Transition in the Frustrated Compound CeMgIn}

\author{Kou Onishi$^{1*}$, Hitoshi Sugawara$^1$, Takahiro Sakurai$^2$, Hitoshi Ohta$^3$, and Eiichi Matsuoka$^1$}
\inst{$^1$Department of Physics, Graduate School of Science, Kobe University, Kobe 675-8501, Japan \\
$^2$Research Facility Center for Science and Technology, Kobe University, Kobe 675-8501, Japan \\
$^3$Molecular Photoscience Research Center, Kobe University, Kobe 675-8501, Japan} 

\abst{We report on the magnetic, transport, and thermal properties of the hexagonal ZrNiAl-type compound \mbox{CeMgIn} with Ce atoms forming a distorted kagome network. This compound exhibits successive antiferromagnetic transition at $T_\text{N1} = \qty{2.1}{\kelvin}$, $T_\text{N2} = \qty{1.7}{\kelvin}$, and possibly $T_\text{N3} = \qty{1.3}{\kelvin}$. The electrical resistivity exhibits a minimum at \qty{11}{\kelvin} and a nonlogarithmic increase with decreasing temperature down to $T_\text{N2}$. We found that \mbox{CeMgIn} is the first ZrNiAl-type compound whose resistivity increase can be well explained by considering a model in which the electron-spin scattering is enhanced by the magnetic frustration and the Ruderman--Kittel--Kasuya--Yosida interaction. These results suggest that \mbox{CeMgIn} is a notable compound whose physical properties are strongly affected by geometrical frustration. Since the Sommerfeld coefficient is \qty{97}{m\joule/\mole \, \kelvin^{2}}, \mbox{CeMgIn} is classified as a moderate heavy-fermion compound.}

\begin{document}
\maketitle

\section{Introduction}
$f$-Electron compounds show a variety of magnetic ground states owing to the competition between the Ruderman--Kittel--Kasuya--Yosida (RKKY) interaction and the Kondo effect.
The magnetic ground states resulting from the competition are described by the Doniach phase diagram\cite{Doniach}.
According to this diagram, magnetic transition due to localized magnetic moments occurs if the RKKY interaction is dominant, whereas nonmagnetic Fermi liquid states are realized if the Kondo effect is dominant. 
Recently, magnetic frustration, which has been considered an important factor for the quantum spin liquid\cite{spin_liqid} or spin ice\cite{spin_ice} in $d$-electron compounds, has been attracting attention as the third key component for the magnetic ground states of $f$-electron compounds; the competition among the magnetic frustration, the RKKY interaction, and the Kondo effect could lead to novel physical properties.
In the theoretical approach, the $Q$--$K$ phase diagram has been proposed to describe the magnetic ground states of frustrated $f$-electron compounds\cite{QK}.
This is a two-axis diagram describing the joint effects of the Kondo screening ($K$) and the quantum zero-point motion induced by frustration ($Q$).
Experimentally, the equiatomic ternary compounds RXX' (R = rare earths; X, X' = {\itshape{d}}- or {\itshape{p}}-block elements), which have the hexagonal ZrNiAl-type crystal structure (space group $P\bar{6}2m$) with no inversion symmetry, have been known to show physical properties affected by the magnetic frustration.
The magnetic frustration in RXX' is attributed to the geometrical frustration due to the distorted kagome network formed by the R atoms.
For example, the antiferromagnetic (AFM) ordered state constructed by 2/3 of $\text{Ce}^{3+}$ ions in CePdAl\cite{CePdAl1}, the kagome spin ice state in HoAgGe\cite{HoAgGe}, the quantum criticality driven by geometrical frustration in CeRhSn\cite{CeRhSn}, and the quantum bicriticality in YbAgGe\cite{YbAgGe} are the notable physical properties affected by magnetic frustration.

\begin{figure}[t]
  \vspace{5truemm}
  \centering
  \iffigure
  \includegraphics[width=0.82\linewidth]{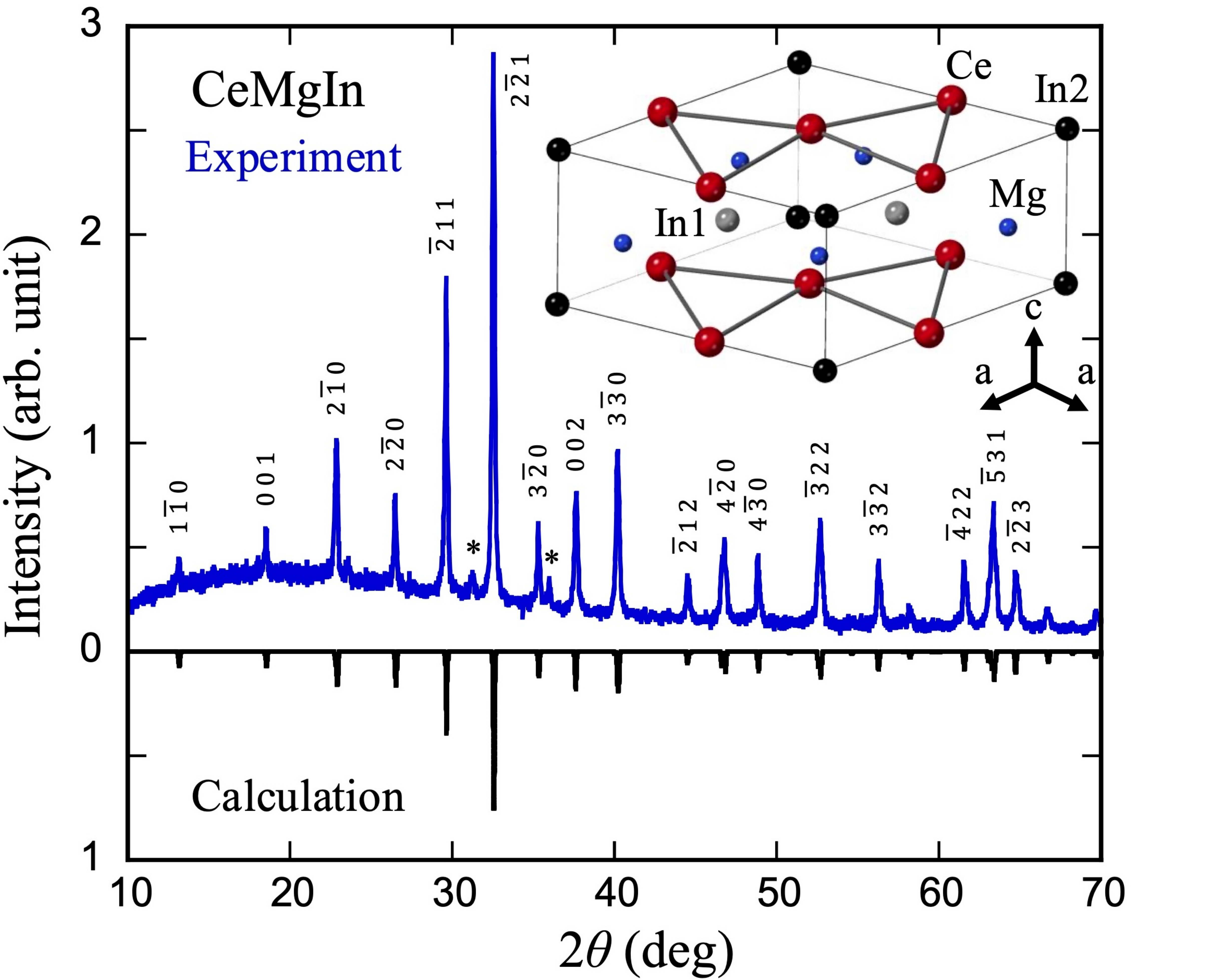}
  \vspace{2.9mm}
  \caption{(Color online). X-ray powder diffraction patterns of \mbox{CeMgIn}. The upper and lower patterns show the experimental and calculated ones, respectively. The numbers shown in the experimental pattern are the Miller indices. The peaks marked with asterisks originate from unidentified impurities. The inset figure shows the chemical unit cell of \mbox{CeMgIn}.}
  \vspace{6.42truemm}
  \label{Fig1}
\end{figure}

To search for the other $f$-electron compounds exhibiting such phenomena, we have focused on the magnesium-containing RXX' compound \mbox{CeMgIn} whose crystal structure is shown in the inset of Fig. \ref{Fig1}.
In contrast to the aforementioned ZrNiAl-type compounds, \mbox{CeMgIn} does not contain transition metals, and thus its conduction band is not affected by the $d$-electron.
In this case, the band structure of \mbox{CeMgIn} could be simpler than that of previously investigated compounds.
Therefore, we expect that the physical properties of \mbox{CeMgIn} would be easier to compare with the theoretical concepts of frustrated systems such as the $Q$--$K$ phase diagram than those of the other ZrNiAl-type compounds, and such comparison will enables us to verify the generality of these concepts.
This compound was synthesized in 2004 as a member of the \mbox{RMgIn} series\cite{RMgIn}.
The magnetic properties of the RMgIn series have been reported for compounds with the heavy rare-earths thus far, i.e., \mbox{DyMgIn} (N\'eel temperature $T_\text{N} = \qty{22}{\kelvin}$), HoMgIn ($T_\text{N} = \qty{12}{\kelvin}$), and TmMgIn ($T_\text{N} = \qty{3}{\kelvin}$) exhibit AFM transition\cite{RMgIn}.
In this paper, we show the results of the magnetic, transport, and thermal properties measurements of \mbox{CeMgIn}.
These results suggest that the physical properties of \mbox{CeMgIn} are strongly affected by magnetic frustration.

\section{Experimental Details}
Polycrystalline samples of \mbox{CeMgIn} and the nonmagnetic reference compound \mbox{LaMgIn} were synthesized by referring to the method reported in Ref. 9.
The ingots of Ce or La (\qty{99.9}{\percent}), Mg rods (\qty{99.99}{\percent}), and In shots (\qty{99.9999}{\percent}) were weighed in the ideal molar ratio of 1:1:1 and sealed in tantalum tubes (length = 40 mm, diameter = 10 mm) under an Ar gas atmosphere.
The tantalum tubes were further sealed in evacuated quartz tubes to avoid oxidization during heating.
The raw elements were melted in a muffle furnace at \qty{1320}{\kelvin} for 10 h, then the mixed elements were annealed at \qty{920}{\kelvin} for one week to ensure homogeneity.
These samples were characterized by X-ray powder diffraction experiments using a diffractometer with Cu-$K_{\alpha1}$ radiation (Rigaku, MiniFlex II).
Figure \ref{Fig1} shows the X-ray diffraction patterns of \mbox{CeMgIn}.
Almost all the Bragg peaks in the experimental patterns can be indexed on the basis of the ZrNiAl-type structure.
The peaks marked with asterisks originate from unidentified impurities.
The intensity of these impurity peaks is about 3\% of that of the $2\, \bar{2}\, 1$ peak, indicating that the volume fraction of the impurities is of similar percentage.
The lattice constants obtained from the X-ray diffraction at room temperature are $a = \qty{7.753(2)}{\angstrom}$ and $c = \qty{4.776(3)}{\angstrom}$ for \mbox{CeMgIn}, and $a = \qty{7.817(3)}{\angstrom}$ and $c = \qty{4.811(4)}{\angstrom}$ for \mbox{LaMgIn}.
These values agree with those reported previously within the range of experimental precision\cite{RMgIn}.
These samples have been stored in evacuated glass tubes to prevent oxidization.
The magnetic, transport, and thermal properties of these samples were examined by magnetization $M$($T$, $H$), electrical resistivity $\rho$($T$, $H$), and specific heat $C$($T$) measurements as functions of temperature $T$ and magnetic field $H$.
$M$($T$, $H$) was measured using a superconducting quantum interference device magnetometer (Quantum Design, MPMS) between 1.8 and \qty{300}{\kelvin} up to \qty{5}{\tesla}.
$\rho$($T$, $H$) was measured by a DC four-terminal method between 0.4 and \qty{300}{\kelvin} up to \qty{9}{\tesla} in a laboratory-built ${}^3$He cryostat.
$C$($T$) was measured by a thermal relaxation method between 0.6 and \qty{9}{\kelvin} in a laboratory-built ${}^3$He cryostat.

\section{Results and Discussion}
\begin{figure}[t]
  \vspace{5truemm}
  \centering
  \iffigure
  \includegraphics[width=0.88\linewidth]{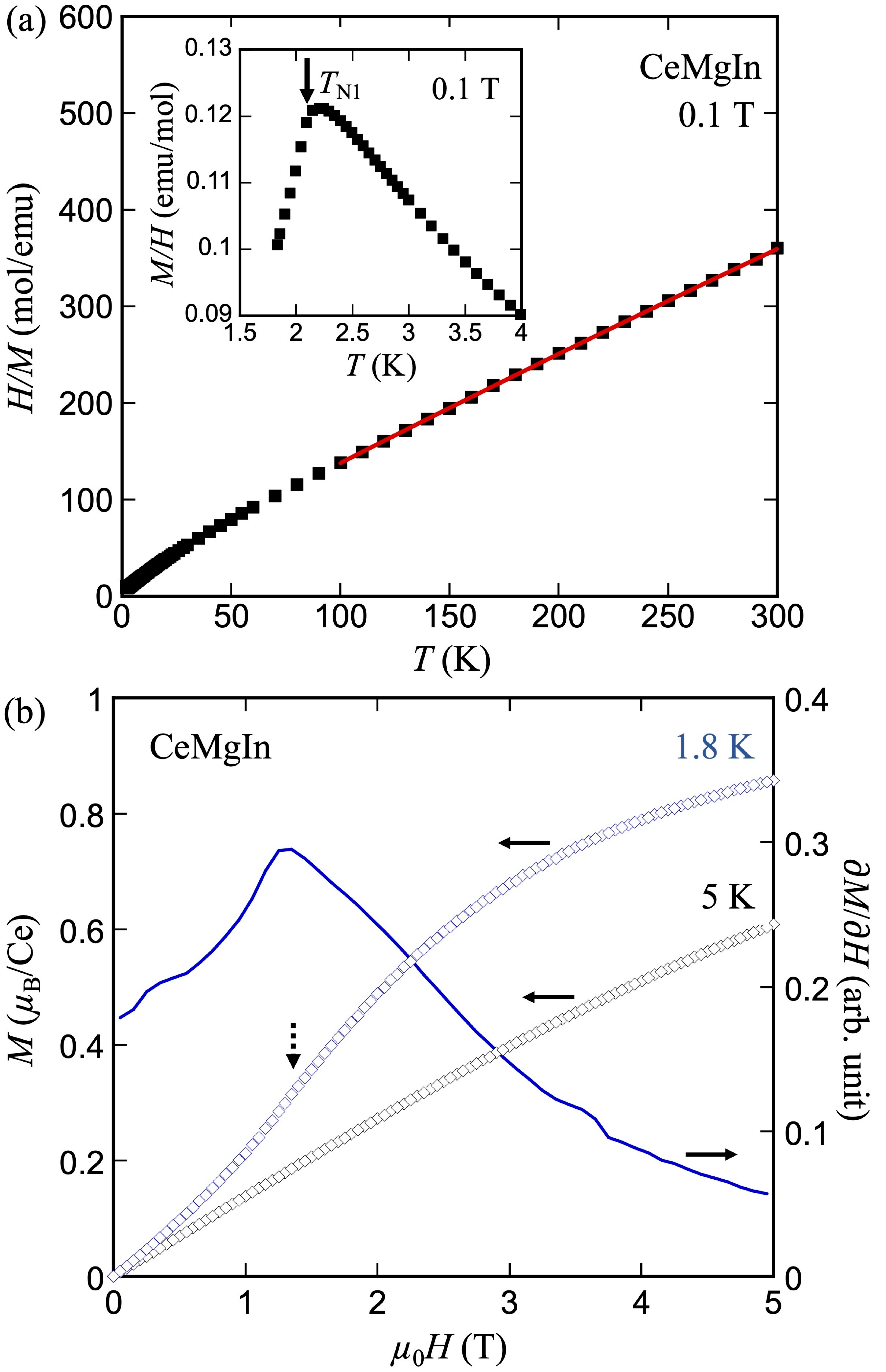}
  \vspace{2.5truemm}
  \caption{(Color online). (a) Temperature dependence of the inverse magnetic susceptibility $H/M(T)$ (main panel) and the magnetic susceptibility $M(T)/H$ (inset) of \mbox{CeMgIn}. The red line is the fitted result using the modified Curie--Weiss law (see text). (b) Magnetic field dependence of the magnetization $M(H)$ at \qty{5}{\kelvin} ($T_\text{N1} < T$) and \qty{1.8}{\kelvin} ($T< T_\text{N1}$) of \mbox{CeMgIn}. The field derivative $\partial M/\partial H$ at 1.8 K is also shown. The dotted arrow at \qty{1.3}{\tesla} (maximum field of $\partial M/\partial H$) represents the magnetic field where spin-flop occurs.}
  \label{Fig2}
  \vspace{4.75truemm}
\end{figure}

Figure \ref{Fig2}(a) shows the temperature dependence of the inverse magnetic susceptibility $H/M(T)$ of \mbox{CeMgIn} measured at \qty{0.1}{\tesla}.
The $H/M(T)$ data above \qty{100}{\kelvin} follows the modified Curie--Weiss law $H/M = \{C_\text{Curie}/(T-\theta_\text{p}) + \chi_0\}^{-1}$, as shown by the red line, where $C_\text{Curie} = \qty{0.827}{emu\,\kelvin/\mole}$, $\theta_\text{p} = \qty{-16.6}{\kelvin}$, and $\chi_0 = 1.7~\times~10^{-4}\, \unit{emu/\mole}$ represent the Curie constant, the paramagnetic Curie temperature, and the sum of the contributions of the Pauli paramagnetism of conduction electrons and the core diamagnetism, respectively.
The effective magnetic moment $\mu_\text{eff}$ calculated from $C_\text{Curie}$ is $2.57\, \mu_\text{B}/\text{Ce}$, suggesting that the Ce ions are trivalent because the $\mu_\text{eff}$ value is close to that of $\text{Ce}^{3+}$ ($2.54\, \mu_\text{B}/\text{Ce}$).
The value of $\chi_0$ is comparable to the $M(T)/H$ value of \mbox{LaMgIn} at \qty{300}{\kelvin} ($1.6 \times 10^{-4}\, \unit{emu/\mole}$).
The inset of Fig. \ref{Fig2}(a) shows the temperature dependence of the magnetic susceptibility $M(T)/H$ below 4 K.
$M(T)/H$ drops below $T_\text{N1} = \qty{2.1}{\kelvin}$, indicating that AFM transition occurs at this temperature.
The frustration parameter $F = |\theta_\text{p}|/T_\text{N}$ is an empirical parameter that represents the degree of suppression of $T_\text{N}$ owing to frustration, and $F > 5 - 10$ indicates that $T_\text{N}$ is strongly suppressed by frustration\cite{spin_liqid}.
In the case of \mbox{CeMgIn}, $F = |\theta_\text{p}|/T_\text{N1} = 7.9$ suggests that $T_\text{N1}$ is moderately suppressed by frustration.
Figure \ref{Fig2}(b) shows the magnetic field dependence of the magnetization $M(H)$ of \mbox{CeMgIn}.
While $M(H)$ at \qty{5}{\kelvin} ($> T_\text{N1}$) exhibits a paramagnetic increase with increasing magnetic field, $M(H)$ at \qty{1.8}{\kelvin} ($< T_\text{N1}$) exhibits a metamagnetic increase at \qty{1.3}{\tesla} as shown by the dotted arrow.
The field of metamagnetic increase is defined as the maximum field of $\partial M/\partial H$.
Since the AFM order occurs below $T_\text{N1}$, the metamagnetic increase can be ascribable to spin-flop.
\begin{figure}[t]
  \vspace{4truemm}
  \centering
  \iffigure
  \includegraphics[width=0.873\linewidth]{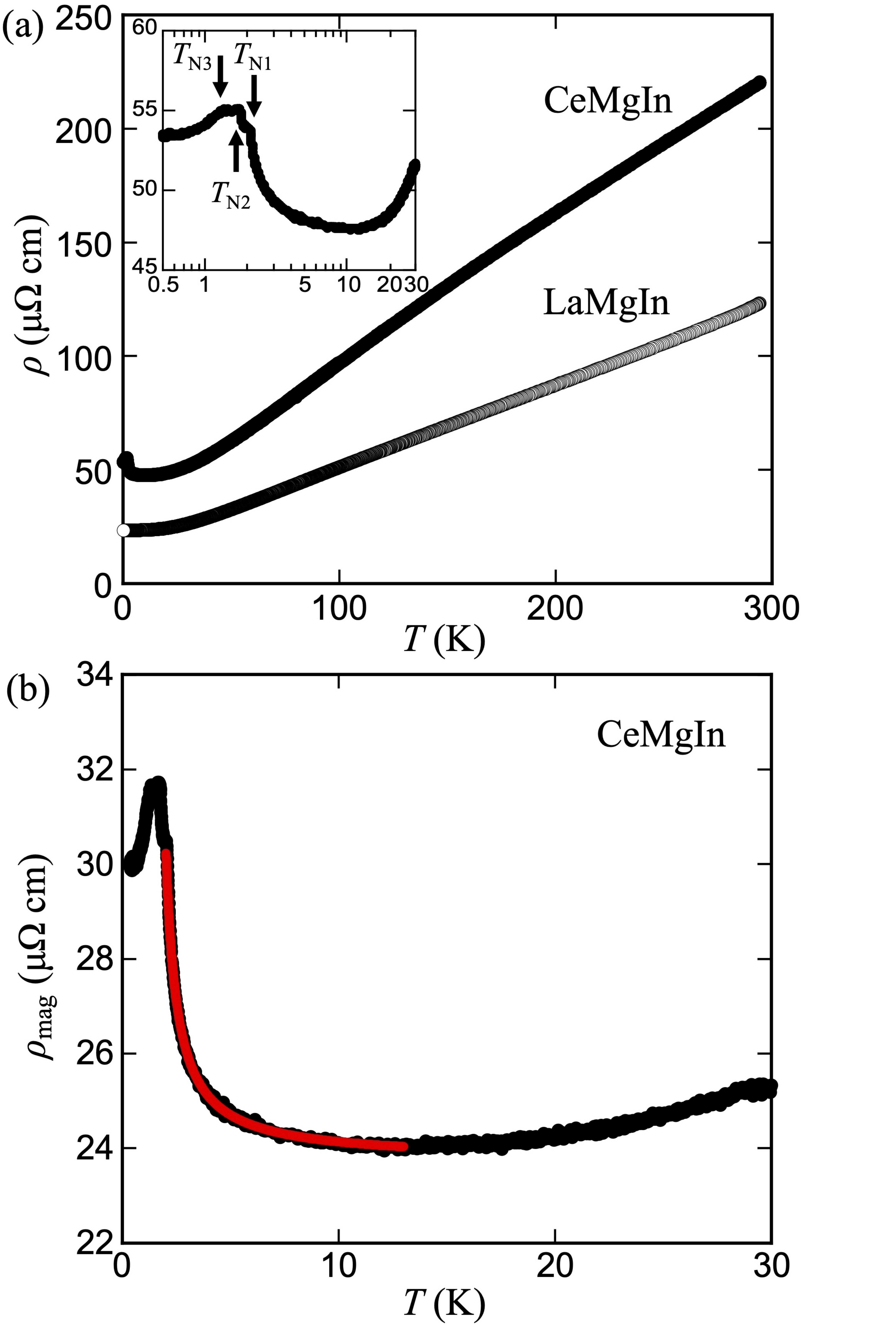}
  \vspace{3.1truemm}
  \caption{(Color online). (a) Temperature dependence of the electrical resistivity $\rho(T)$ of \mbox{CeMgIn}. $\rho(T)$ of \mbox{LaMgIn} is shown for comparison. The inset shows the low temperature part of the temperature dependence of $\rho(T)$ of \mbox{CeMgIn}. (b) Temperature dependence of $\rho_\text{mag}(T)$ estimated by subtracting $\rho(T)$ of \mbox{LaMgIn} from that of \mbox{CeMgIn}. The red line represents the calculation result considering the electron-spin scattering enhanced by the RKKY interaction (see text).}
  \label{Fig3}
  \vspace{5.8truemm}
\end{figure}

Figure \ref{Fig3}(a) shows the temperature dependence of the electrical resistivities $\rho(T)$ of \mbox{CeMgIn} and the nonmagnetic reference compound \mbox{LaMgIn}.
$\rho(T)$ of \mbox{CeMgIn} exhibits a metallic decrease from room temperature, followed by a minimum at \qty{11}{\kelvin}, as shown in the inset, then a steep increase below the minimum temperature.
The possible origins of the minimum and the steep increase in $\rho(T)$ will be discussed later.
$\rho(T)$ subsequently exhibits an inflection at $T_\text{N1}$, and continues to increase down to $T_\text{N2} = \qty{1.7}{\kelvin}$.
The increase in $\rho(T)$ between $T_\text{N1}$ and $T_\text{N2}$ suggests that the contribution of $\rho(T)$ increase due to the formation of a superzone gap overcomes that of $\rho(T)$ decrease due to the reduction of magnetic scattering.
$\rho(T)$ then decreases at $T_\text{N2}$ and $T_\text{N3} = \qty{1.3}{\kelvin}$, suggesting that some phase transition occurs at these temperatures.

\begin{figure}[t]
  \vspace{4truemm}
  \centering
  \iffigure
  \includegraphics[width=0.867\linewidth]{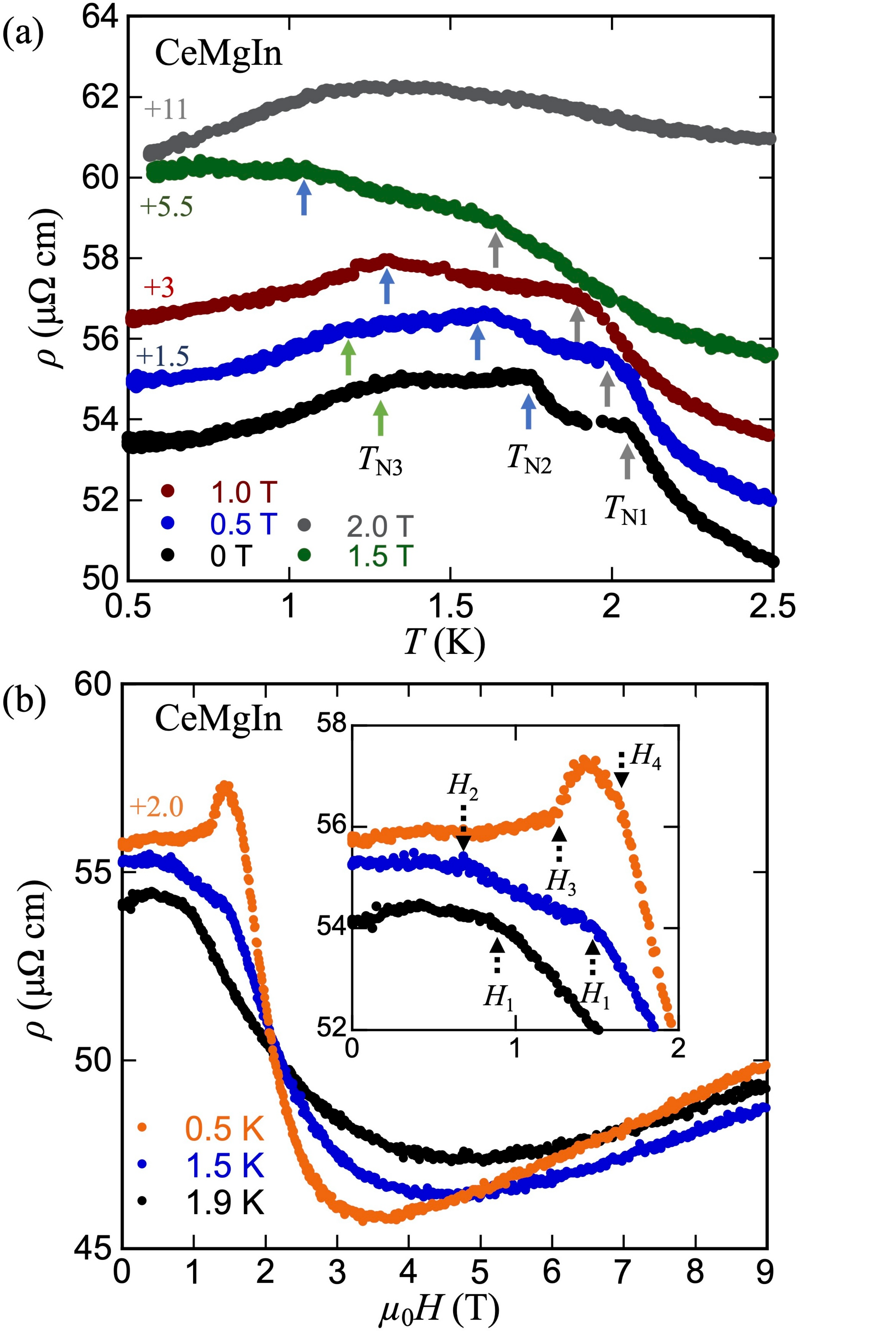}
  \vspace{3.1truemm}
  \caption{(Color online). (a) Temperature dependence of $\rho(T)$ in various magnetic fields of \mbox{CeMgIn}. The values of $\rho(T)$ are vertically offset by the values shown in the figure for easy observation. (b) Magnetic field dependence of resistivity $\rho(H)$ (main panel) and the low field part (inset) of \mbox{CeMgIn}. $H_1$, $H_2$, $H_3$, and $H_4$ correspond to the magnetic fields of phase boundaries. The value of $\rho(H)$ at \qty{0.5}{\kelvin} is vertically offset by $\qty{2.0}{\micro\ohm \, c\meter}$ for easy observation.}
  \label{Fig4}
  \vspace{5.82truemm}
\end{figure}

We then discuss the possible origins of the minimum and the steep increase in $\rho(T)$.
The resistivity minimum of Ce compounds is often observed and attributed to the Kondo effect.
In that case, $\rho(T)$ shows $-\ln{T}$ dependence below the minimum temperature.
However, in the case of \mbox{CeMgIn}, the nonlogarithmic increase in $\rho(T)$ is evident, as shown in the inset of Fig. \ref{Fig3}(a).
This result suggests that the Kondo effect alone is insufficient to explain the $\rho(T)$ increase.
The other possible origin of the resistivity increase is the electron-spin scattering enhanced by the RKKY interaction in frustrated magnets\cite{R_mini}.
A strong frustration stabilizes a liquid-like state of localized magnetic moments down to temperatures well below $|\theta_\text{p}|$.
In this circumstance, the RKKY interaction enhances the elastic scattering of conduction electrons, resulting in the resistivity upturn below $|\theta_\text{p}|$\cite{R_mini}.
In fact, the resistivity increases in the frustrated compounds SmCuAs${}_2$\cite{RCuAs2} and GdCuIn${}_4$\cite{RCuIn4} have been explained by considering this mechanism\cite{R_mini2}.
In the case of \mbox{CeMgIn}, the magnetic contribution to the resistivity $\rho_\text{mag}(T)$ estimated by subtracting $\rho(T)$ of \mbox{LaMgIn} from that of \mbox{CeMgIn} starts to increase below \qty{13}{\kelvin}, which is close to $|\theta_\text{p}|$ = 16.6 K, and continues to increase down to $T_\text{N1}$, as shown in Fig. \ref{Fig3}(b).
Considering the mechanism discussed in Ref. 10, we expect that the geometrical frustration on the distorted kagome network develops a liquid-like state of the Ce moments between $T_\text{N1}$ and 13 K, and the  RKKY interaction in this state enhances the magnetic scattering of conduction electrons, resulting in the resistivity increase in \mbox{CeMgIn}.
Hereafter, we analyze $\rho(T)$ of \mbox{CeMgIn} by referring to the discussion in Refs. 10 and 13.
The temperature dependence of the resistivity due to the electron-spin scattering is expressed as $\rho_\text{RKKY}(T) = a/(T - T^*) + b$\cite{R_mini, R_mini2}.
The exact formulae for $a$, $b$, and $T^*$ are shown in Ref. 13.
The $\rho_\text{mag}(T)$ between 2.1 and \qty{13}{\kelvin} is well explained by the above formula with $a = \qty{3.37}{\micro\ohm \, c\meter \, \kelvin}$, $b = \qty{23.7}{\micro\ohm \, c\meter}$, and $T^* = \qty{1.50}{\kelvin}$, as shown by the red line.
Thus, the contribution of electron-spin scattering due to the RKKY interaction is dominant in the increase in $\rho(T)$ with decreasing temperature, rather than the contribution of the Kondo effect.
Therefore, we conclude that magnetic frustration plays a crucial role in the low-temperature $\rho(T)$ behavior of \mbox{CeMgIn}.

Here, we compare the $\rho(T)$ increase of \mbox{CeMgIn} with that of the isostructural compound CePdAl.
$\rho(T)$ of CePdAl exhibits a minimum at around \qty{20}{\kelvin} and $-\ln{T}$ dependence down to \qty{4}{\kelvin}\cite{CePdAl2}.
Since the Kondo temperature of \mbox{CeMgIn} is comparable to that of CePdAl, as described later, one might conjecture that $\rho(T)$ of \mbox{CeMgIn} exhibits $-\ln{T}$ dependence owing to the Kondo effect.
However, this conjecture contradicts our experimental results.
We now speculate that the difference in the $\rho(T)$ increase between \mbox{CeMgIn} and CePdAl is attributed to the band structures of the conduction electrons, which are responsible for both the formation of the Kondo singlet and the mediation of the RKKY interaction.
Although the dominant contribution of the RKKY interaction might be a result of the simplicity of the band structure arising from the lack of $d$-electrons, our current results do not provide any explicit information about the band structure of \mbox{CeMgIn}.
Hence, the band structure of \mbox{CeMgIn} should be examined experimentally and theoretically, and it should be compared with that of CePdAl to investigate the difference in the factors responsible for the $\rho(T)$ increase.

Figure \ref{Fig4}(a) shows the low-temperature part in $\rho(T)$ of \mbox{CeMgIn} measured in various magnetic fields.
With increasing magnetic field up to \qty{1.5}{\tesla}, $T_\text{N1}$, $T_\text{N2}$, and $T_\text{N3}$ shift to lower temperature.
This finding suggests that the phase transitions at these temperatures are antiferromagnetic.
No anomalies have been observed above \qty{2.0}{\tesla}.
Figure \ref{Fig4}(b) shows the magnetic field dependence of the resistivity $\rho(H)$ of \mbox{CeMgIn} measured at various temperatures.
$\rho(H)$ at \qty{1.9}{\kelvin} exhibits one shoulder at $H_1 = \qty{0.9}{\tesla}$, while $\rho(H)$ at \qty{1.5}{\kelvin} exhibits two shoulders at $H_2 = \qty{0.7}{\tesla}$ and $H_1 = \qty{1.5}{\tesla}$, as shown in the inset of Fig. \ref{Fig4}(b).
$\rho(H)$ at \qty{0.5}{\kelvin} increases rapidly above $H_3 = \qty{1.2}{\tesla}$ and exhibits a shoulder at $H_4 = \qty{1.6}{\tesla}$.
$H_1$, $H_2$, $H_3$, and $H_4$ correspond to the magnetic fields of the phase boundaries, as described later.

\begin{figure}[t]
  \vspace{4truemm}
  \centering
  \iffigure
  \includegraphics[width=0.822\linewidth]{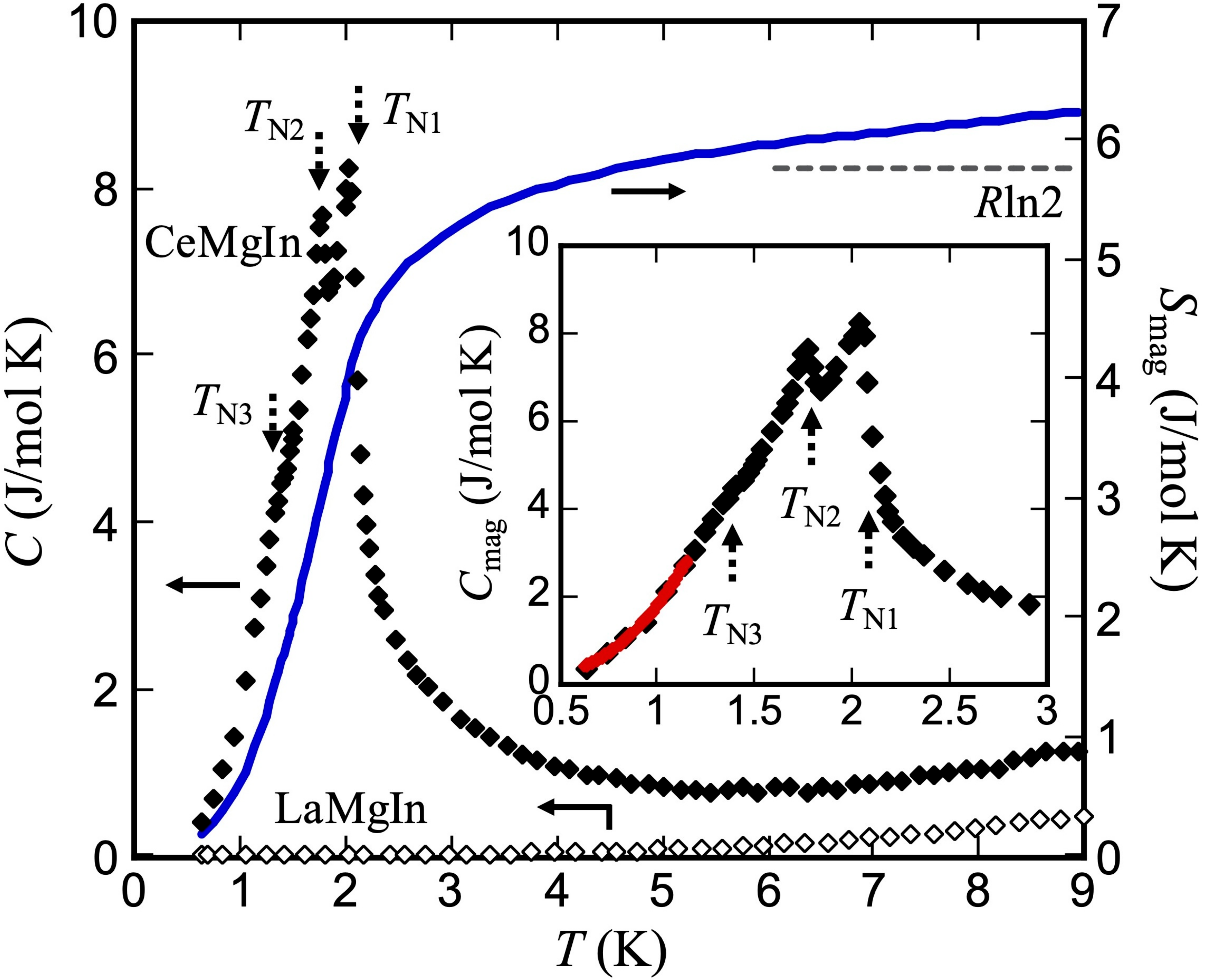}
  \vspace{2.8truemm}
  \caption{(Color online). Temperature dependences of the specific heat $C(T)$ and the magnetic entropy $S_\text{mag}(T)$ of \mbox{CeMgIn}. $C(T)$ of \mbox{LaMgIn} is shown for comparison. The inset shows the temperature dependence of the magnetic specific heat $C_\text{mag}(T)$ obtained by subtracting the $C(T)$ of \mbox{LaMgIn} from that of \mbox{CeMgIn}. The red line represents the calculation result considering the electronic specific heat and the gapped AFM magnon (see text).}
  \label{Fig5}
  \vspace{4.8truemm}
\end{figure}

Figure \ref{Fig5} shows the temperature dependence of the specific heats $C(T)$ of \mbox{CeMgIn} and \mbox{LaMgIn}.
$C(T)$ of \mbox{CeMgIn} exhibits a $\lambda$-type peak at $T_\text{N1}$, suggesting that the second-order AFM transition occurs at this temperature.
$C(T)$ also exhibits a sharp peak and a broad shoulder at $T_\text{N2}$ and $T_\text{N3}$, respectively.
The sharp $C(T)$ peak corroborates that the anomaly of $\rho(T)$ at $T_\text{N2}$ is attributed to the intrinsic AFM transition.
In contrast, one might suspect that the broad $C(T)$ shoulder at $T_\text{N3}$ is due to the phase transition of the unidentified impurities detected by X-ray diffraction.
Although no \mbox{Ce-Mg-In} compounds have been reported to exhibit the phase transition at $T_\text{N3}$, we cannot completely rule out the possibility that the broad shoulder of $C(T)$ and the $\rho(T)$ decrease at $T_\text{N3}$ are attributed to impurities.
We need to verify whether or not the phase transition at $T_\text{N3}$ is intrinsic by further experiments in the future.
The inset of Fig. \ref{Fig5} shows the temperature dependence of the magnetic specific heat $C_\text{mag}(T)$ estimated by subtracting $C(T)$ of \mbox{LaMgIn} from that of \mbox{CeMgIn}.
The $C_\text{mag}(T)$ data below \qty{1.2}{\kelvin} can be fitted by the formula of $C_\text{mag} = \gamma T + \beta T^3 \exp{(-\Delta/k_\text{B}T)}$, as shown by the red line, where the first and the second terms represent the electronic specific heat and the contribution of the gapped AFM magnon, respectively.
Here, the obtained \nobreak{Sommerfeld} coefficient $\gamma = \qty{97}{m\joule/\mole \, \kelvin^{2}}$ is about 186 times larger than that of \mbox{LaMgIn} (\qty{0.52}{m\joule/\mole \, \kelvin^{2}}), suggesting that \mbox{CeMgIn} is a heavy fermion compound with moderately enhanced effective mass.
The values of $\beta$ and the energy gap of the magnon, $\Delta/k_\text{B}$, are $2.4\,\unit{\joule/\mole \, \kelvin^{4}}$ and $\qty{0.35}{\kelvin}$, respectively.
The fitting result for $C_\text{mag}(T)$ considering the gapped magnon implies an anisotropic arrangement of Ce moments. 
The temperature dependence of the magnetic entropy $S_\text{mag}(T)$ of \mbox{CeMgIn} was calculated by integrating $C_\text{mag}/T$ with $T$.
Since the local symmetry of the $\text{Ce}^{3+}$ sites is orthorhombic, the sixfold ground multiplet of the $J = 5/2$ state splits into three Kramers doublets.
One can therefore expect that the magnetic entropy of $R\ln{2}$ ($R$: gas constant) corresponding to a ground Kramers doublet is released at $T_\text{N1}$.
However, $S_\text{mag}(T)$ at $T_\text{N1}$ is $\qty{4.28}{\joule/\mole \, \kelvin}$, which is 74\% of $R\ln{2}$.
There are two possible reasons for the suppression of $S_\text{mag}(T)$ from $R\ln{2}$: the development of short-range magnetic correlations above $T_\text{N1}$ and the shielding of magnetic moments due to the Kondo effect.
The former is corroborated by the gradual increase in $C(T)$ with decreasing temperature below \qty{5}{\kelvin}.
The liquid-like state of localized moments described above should be responsible for the development of short-range correlations.
The latter effect is evaluated by considering the magnitude of the Kondo temperature $T_\text{K}$.
We obtain $T_\text{K} = \qty{6.5}{\kelvin}$ by analyzing the $S_\text{mag}(T)$ data using a two-level model with an energy splitting $k_\text{B}T_\text{K}$\cite{Kondo.Temp}.
This value is comparable to that of the isostructural Kondo lattice compound CePdAl ($T_\text{K}\, = \qty{5}{\kelvin}$)\cite{CePdAl_kondo}, whose $S_\text{mag}$ is strongly reduced by the Kondo effect\cite{CePdAl_capacity}, suggesting that the Kondo effect contributes to the suppression of $S_\text{mag}(T)$ in \mbox{CeMgIn}.
The fact that the value of $S_\text{mag}$ exceeds $R\ln{2}$ above 5 K suggests that the first excited state is a few tens of Kelvin away from the ground state.

\begin{figure}[t]
  \vspace{4truemm}
  \centering
  \iffigure
  \includegraphics[width=0.845\linewidth]{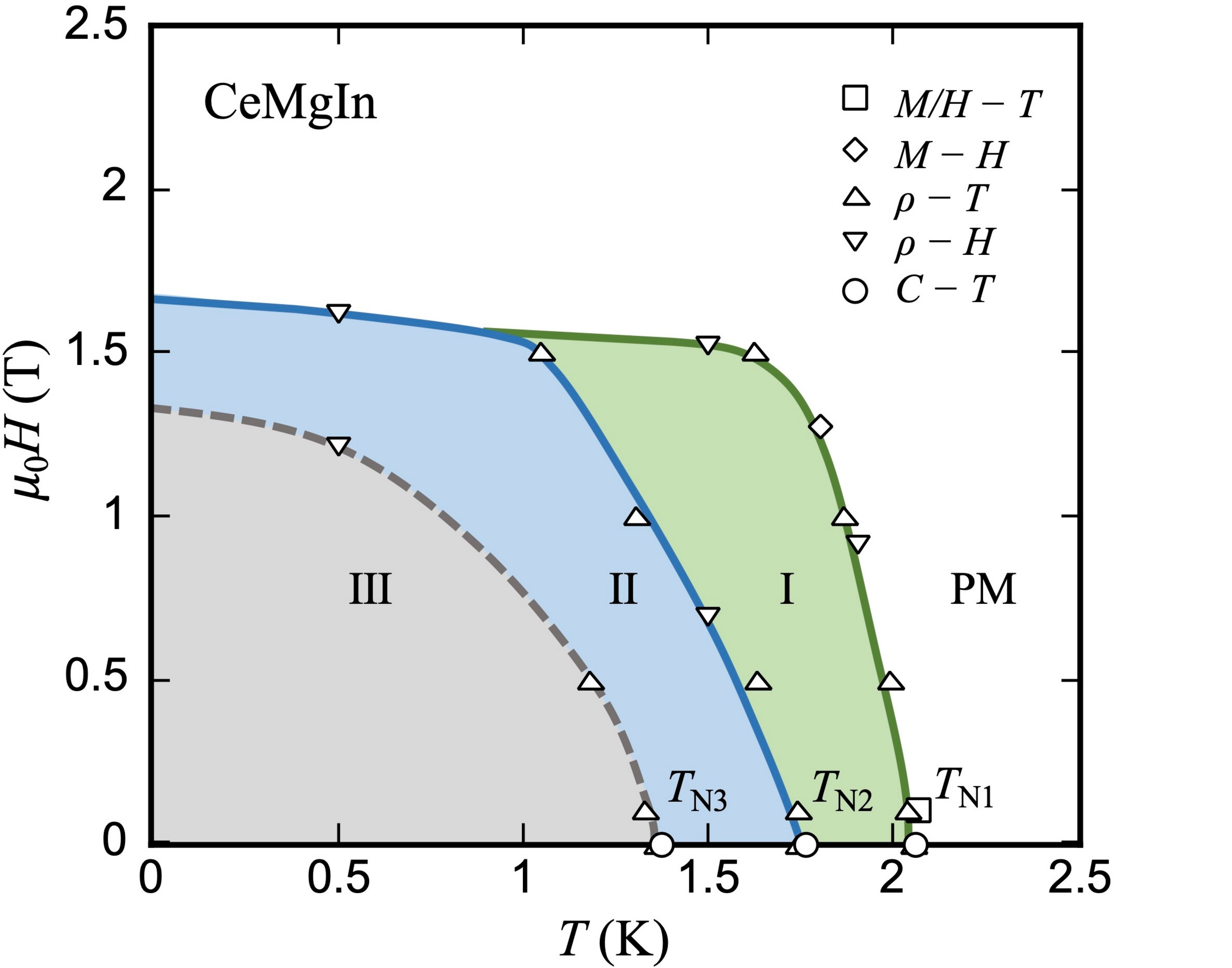}
  \vspace{2.3truemm}
  \caption{(Color online). Magnetic field $H$ vs temperature $T$ phase diagram of \mbox{CeMgIn}. Transition temperatures and magnetic fields determined by $M(T)/H$, $M(H)$, $\rho(T, H)$, and $C(T)$ measurements are represented by a square, diamonds, triangles, and circles, respectively. PM represents a paramagnetic phase and I, I\hspace{-1pt}I, and I\hspace{-1pt}I\hspace{-1pt}I represent AFM phases.}
  \label{CeMgIn_TB}
  \vspace{8.41truemm}
\end{figure}

Figure \ref{CeMgIn_TB} shows the $H$--$T$ phase diagram constructed from the measurements of $M(T)/H$, $M(H)$, $\rho(T, H)$, and $C(T)$.
PM represents the paramagnetic phase and I, I\hspace{-1pt}I, and I\hspace{-1pt}I\hspace{-1pt}I represent AFM phases.
$H_1$, $H_2$, $H_3$, and $H_4$ shown in the inset of Fig.~\ref{Fig4}(b) correspond to the magnetic fields of the phase boundaries between PM and I, between I and I\hspace{-1pt}I, between I\hspace{-1pt}I and I\hspace{-1pt}I\hspace{-1pt}I, and between PM and I\hspace{-1pt}I, respectively.
Since we cannot rule out the possibility that the phase transition at $T_\text{N3}$ is caused by impurities, the phase boundary between \mbox{I\hspace{-1pt}I} and \mbox{I\hspace{-1pt}I\hspace{-1pt}I} is represented by a dotted line.
The most characteristic feature of this diagram is the existence of the multiple AFM phases. 
The multiple AFM phases (or the successive AFM transition) were also observed in the isostructural frustrated compounds PrPdAl\cite{PrPdAl_neutron, PrPdAl_property} and HoAgGe\cite{RAgGe}.
Particularly in HoAgGe, the successive phase transition from an AFM structure contributed by 2/3 of the $\text{Ho}^{3+}$ ions and fluctuating 1/3 of the $\text{Ho}^{3+}$ ions, to an AFM structure contributed by all of the $\text{Ho}^{3+}$ ions is attributed to the rearrangement of magnetic moments within the $c$-plane\cite{HoAgGe}.
In the case of \mbox{CeMgIn}, the successive AFM transition could also be attributed to the successive rearrangement of magnetic moments.
To verify this conjecture, the magnetic structures of I, \mbox{I\hspace{-1pt}I}, and I\hspace{-1pt}I\hspace{-1pt}I phases should be determined by neutron diffraction experiments.

\section{Summary}
In this study, we prepared polycrystalline samples of \mbox{CeMgIn} and revealed the magnetic, transport, and thermal properties.
We discovered that \mbox{CeMgIn} exhibits the successive AFM transition at $T_\text{N1} = \qty{2.1}{\kelvin}$, $T_\text{N2} = \qty{1.7}{\kelvin}$, and possibly $T_\text{N3} = \qty{1.3}{\kelvin}$.
The frustration parameter $F = 7.9$ suggests the relatively strong suppression of $T_\text{N1}$ by frustration.
The $\rho_\text{mag}(T)$ increase on cooling from 13 to \qty{2.1}{\kelvin} is well explained by considering a model in which the electron-spin scattering is enhanced by the magnetic frustration and the RKKY interaction.
\mbox{CeMgIn} is the first ZrNiAl-type compound whose resistivity increase is explained by the above model, and it will be interesting to confirm whether the same model can explain the resistivity increase in the other ZrNiAl-type compounds, such as HoAgGe\cite{RAgGe}, as future tasks.
The increase in $\rho(T)$ for $T_\text{N2} < T < T_\text{N1}$ is attributed to the formation of a superzone gap.
Since the $\gamma = \qty{97}{m\joule/\mole \, \kelvin^{2}}$ of \mbox{CeMgIn} is 186 times larger than that of \mbox{LaMgIn}, \mbox{CeMgIn} is classified as a moderate heavy-fermion compound.
To determine the magnetic structure of the AFM phases, neutron powder diffraction experiments are currently scheduled.
Furthermore, we are now preparing single-crystal samples to reveal magnetic anisotropy and to construct a more detailed $H$--$T$ phase diagram and are planning de Haas--van Alphen effect measurements to examine the band structure.
\newpage

\begin{acknowledgment}
\footnotesize{\textbf{Acknowledgements} \,\,\, The synthesis of the samples was partly conducted using the MatNavi provided by the Materials Data Platform (MDPF) of the National Institute for Materials Science (NIMS). This work was supported by JSPS KAKENHI (Grant Numbers 18H04320, 20K03861, and 21K03470), JST SPRING (Grant
Number JPMJSP2148), and the joint research program of the Molecular Photoscience Research Center of Kobe University. \par}
\end{acknowledgment}

\end{document}